\documentstyle{prhep97}
\input epsf

\makeatletter
\let\chapter\hid@chapter
\makeatother
\renewcommand{\thepage}{} 

\authorrunning{T. Blum and A.\,Soni}
\titlerunning{{\talknumber}: Update on lattice QCD with domain wall quarks}
 
\def\talknumber{1505}

\newcommand{\LL}{\left\langle}
\newcommand{\RR}{\right\rangle}

\newcommand{\BE}{\begin{equation}}
\newcommand{\EE}{\end{equation}}
\newcommand{\BEA}{\begin{eqnarray}}
\newcommand{\EEA}{\end{eqnarray}}

\newcommand{\etal}{{\em et al.\ }}

\newcommand{\vs}{{\em vs.\ }}

\newcommand{\gbeta}{6/g^2}

\def\simge{
    \mathrel{\rlap{\raise 0.511ex
        \hbox{$>$}}{\lower 0.511ex \hbox{$\sim$}}}}
\def\simle{
    \mathrel{\rlap{\raise 0.511ex
        \hbox{$<$}}{\lower 0.511ex \hbox{$\sim$}}}}

\newcommand{\AmS}{{\protect\the\textfont2
  A\kern-.1667em\lower.5ex\hbox{M}\kern-.125emS}}

\begin{document}

\title{{\talknumber}:  Update on lattice QCD with domain wall quarks}
\author{Tom\,Blum\inst{1} (tblum@bnl.gov) \\
Amarjit\,Soni\inst{1} (soni@bnl.gov); Presenter}
\institute{Brookhaven National Laboratory, Upton, NY\ \ 11973}

\maketitle

\begin{abstract}
Using domain wall fermions, 
we estimate $B_K(\mu\approx 2\,{\rm GeV})=0.602(38)$ in 
quenched QCD which is consistent with
previous calculations. We also find ratios of decay constants that are
consistent with experiment, within our statistical errors.
Our initial results indicate good scaling behavior
and support expectations that $O(a)$ errors are
exponentially suppressed in low energy ($E\ll a^{-1}$) observables.
It is also shown that the axial current
numerically satisfies the lattice analog of the usual 
continuum axial Ward identity and that the matrix element of the 
four quark operator needed for $B_K$ exhibits excellent chiral behavior.
\end{abstract}

\section{ Introduction }

We recently reported\cite{US,US2} on calculations
using a new discretization for simulations of QCD, domain wall fermions
(DWF)~\cite{KAPLAN,SHAMIR}, which preserve chiral symmetry on the lattice
in the limit of an infinite extra 5th dimension.
There it was demonstrated
that DWF exhibit remarkable chiral behavior\cite{US} even at relatively large
lattice spacing and modest extent of the fifth dimension.
 
In addition to retaining chiral symmetry, DWF are also
``improved'' in another important way. 
In the limit that the number of sites in the extra dimension, $N_s$, goes 
to infinity, the leading discretization error in the
effective four dimensional action for the light degrees of freedom goes like
$O(a^2)$. This theoretical dependence
is deduced from the fact that the only operators available to
cancel $O(a)$ errors in the effective action are not chirally 
symmetric~\cite{NEUB,US2}.
For finite $N_s$, $O(a)$ corrections are expected to be exponentially 
suppressed with the size of the extra fifth dimension. 
Our calculations for $B_K$ show 
a weak dependence on $a$ that is easily fit to an $a^2$ ansatz. Preliminary
results for the ratios $f_\pi/m_\rho$ and $f_k/f_\pi$ 
indicate good scaling behavior as well.
 
We use the boundary fermion variant of DWF developed by Shamir. For details,
consult Kaplan\cite{KAPLAN} and Shamir\cite{SHAMIR}. 
See Ref.~\cite{SHAMIRandFURMAN} for a discussion of the $4d$ chiral 
Ward identities (CWI) satisfied by DWF.
Our simulation parameters are summarized in Table~\ref{RUNTABLE}. 
\begin{table}[hbt]
\caption{Summary of simulation parameters.
$M$ is the five dimensional Dirac
fermion mass, and $m$ is the coupling between layers $s=0$ and 
$N_s-1$.}
\begin{center}
\begin{tabular}{|c|c|c|c|}
\hline
$\gbeta$ & size & $M$ & $m$($\#$ conf)\\
\hline
5.85& $16^3\times 32\times 14$& 1.7& 0.075(34) 0.05(24)\\
\hline
6.0 & $16^3\times 32\times 10$& 1.7& 0.075(36) 0.05(39) 0.025(34)\\ 
\hline
6.3 & $24^3\times 60\times 10$& 1.5& 0.075(11) 0.05(15) 0.025 (22)\\
\hline
\end{tabular}
\label{RUNTABLE}
\end{center}
\end{table}

\section{ Results }

We begin with the numerical investigation
of the lattice PCAC relation.
The CWI are satisfied exactly on any configuration
since they are derived from the corresponding operator identity. We
checked this explicitly in our simulations.
In the asymptotic large time limit, we find for the usual PCAC
relation
\BEA
2\sinh{\left(\frac{a m_\pi}{2}\right)}
\frac{\LL A_\mu|\pi\RR}{\LL J_5|\pi \RR}&=&
2 m +2\frac{\LL J_{5q}|\pi\RR}{\LL J_5|\pi\RR},
\label{ratio}
\EEA
which goes over to the continuum relation for $a m_\pi \ll 1 $ and 
$N_s\to\infty$ (see Ref.~\cite{SHAMIRandFURMAN} for operator definitions).
The second term on the r.h.s. is anomalous and vanishes as
$N_s\to\infty$. It is a measure of explicit chiral symmetry breaking
induced by the finite 5th dimension.
At $\gbeta=6.0$ and $N_s=10$ we find the l.h.s. of Eq.\ref{ratio} to be
0.1578(2) and 0.1083(3) for $m=0.075$ and 0.05, respectively. The 
anomalous contributions for these two masses are $2\times$(0.00385(5) and
0.00408(12)), which appear to be roughly constant with $m$.
Increasing $N_s$ to 14 at $m=0.05$,
the anomalous contribution falls to
$(2\times)$ 0.00152(8) while the l.h.s. is 0.1026(6), 
which shows that increasing $N_s$ really does take us
towards the chiral limit.
\begin{figure}[hbt]
    \hskip .75in\vbox{ \epsfxsize=3.0in \epsfbox[0 0 4096 4096]{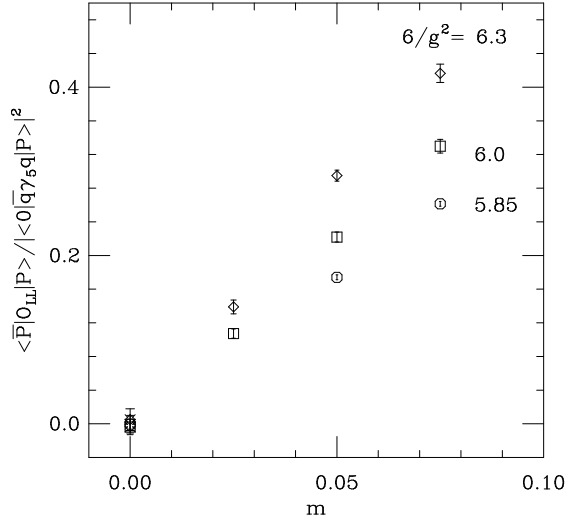} }
    \caption{The matrix element $\LL \bar P |O_{LL}| P \RR$ \vs $m$.
       $m$ is proportional to the quark mass in lattice units. 
       $N_s=10$ ($\gbeta=6.0$, 6.3) and 14 ($\gbeta=5.85$).}
    \label{mll}
\end{figure}

Next we investigate the matrix element of the four quark operator $O_{LL}$ 
which defines $B_K$.
$\LL K|O_{LL}|\bar K\RR$ vanishes linearly
with $m$ in the chiral limit in excellent agreement with chiral perturbation
theory (Fig.~\ref{mll}). 
\begin{figure}[hbt]
   \hskip.75in\vbox{ \epsfxsize=3.0in \epsfbox[0 0 4096 4096]{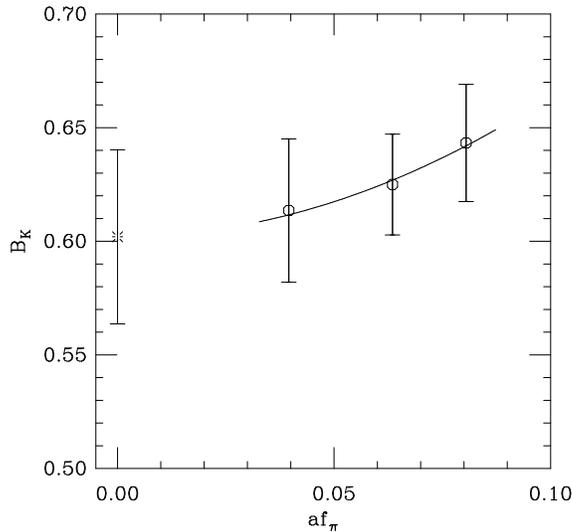} }
    \caption{The kaon B parameter.}
    \label{BK}
\end{figure}

In Fig.~\ref{BK} we show the kaon B parameter at each value of
$\gbeta$ versus $af_\pi$ which is used to set the lattice spacing. 
The results for $B_K$ depend weakly on $\gbeta$, and are well fit to a
pure quadratic in $a$. We find $B_K(\mu=a^{-1})=0.602(38)$
in the continuum limit. This value is already consistent
with previous results\cite{BPAR} though it does not include the
perturbative running of $B_K$ to a common energy scale.
This requires a perturbative calculation to determine the scale dependence
of $O_{LL}$, which has not yet been done\cite{ONELOOP}.
From Table~\ref{decay table}, the energy scale at $\gbeta=6.0$ is roughly 2 GeV.

At $\gbeta=6.0$,
we have also calculated $B_K$ using the partially conserved axial
current $A^a_\mu(x)$ (and the analogous vector current).
This point split conserved current requires explicit factors of the gauge links
to be gauge invariant. Alternatively a gauge non-invariant operator
may be defined by omitting the links; the two definitions
become equivalent in the continuum limit.
Results for the gauge non-invariant operators agree within small 
statistical errors with those obtained with
naive currents, Fig.~\ref{BK}(see Ref.\cite{SHAMIRandFURMAN,US}
for operator definitions). 
The results for the gauge invariant operators
are somewhat larger: $B_K^{inv}(\mu=a^{-1})=0.857(20)$ and 0.946(28)
at $m=0.05$ and 0.075, respectively. A similar situation holds in the
Kogut-Susskind case where it was shown that the gauge invariant operators
receive appreciable perturbative corrections which bring the two results
into agreement~\cite{ISH}.
  
\begin{table}[hbt]
\caption{Lattice spacing and decay constant summary.}
\begin{center}
\begin{tabular}{|c|c|c|c|c|c|}
\hline
$\gbeta$ & $a^{-1}(m_\rho)$ & $a^{-1}(f_\pi)$ & $f_\pi/m_\rho$ & $f_K/f_\pi$  & $N_s$\\
\hline
5.85&  1.49(29)&  1.62(27)&  0.154(40)& 1.206(15)  & 14\\
\hline
6.0 &  1.89(14)&  2.06(15)&  0.155(11)& 1.205(15)  & 10\\
\hline
6.3 &  2.96(25)&  3.24(27)&  0.155(19)& 1.14(14)   & 10\\
\hline
\end{tabular}
\label{decay table}
\end{center}
\end{table}

Using Eq.~\ref{ratio}, neglecting the anomalous contribution, and
using the definition of the decay constant,
we can determine the pseudoscalar decay constant from 
the measurement of $\LL 0| J^a_5|P\RR$.
The results are summarized in Table~\ref{decay table}.  As with
$B_K$, the estimates of the physical ratios $f_\pi/m_\rho$ and $f_K/f_\pi$
indicate good scaling behavior. They are also consistent
with experiment, within rather large statistical errors. 
The errors in Table~\ref{decay table} are crude 
estimates derived using the statistical uncertainties only; 
the small sample size precludes us from properly accounting
for the correlations in the data. All of the results are so called
effective values, calculated from the two-point correlators without
$\chi^2$ fits and averaged over suitable plateaus. 
In the last column at the 
two larger couplings, an attempt has been made to carry out a more 
sophisticated jackknife error analysis all the way through to the
final ratio.
  
While all of the above results indicate good scaling, they
must be checked further with improved statistics and a fully
covariant fitting procedure. Also systematic effects like
finite volume still need to be investigated. Only then 
can the continuum limit can be reliably taken.
We note that a recent precise calculation using quenched Wilson
quarks by the CP-PACS collaboration yields values of $f_\pi$ and $f_K$ in
the continuum limit that are inconsistent with experiment~\cite{CPPACS}.
\begin{figure}[hbt]
    \hskip .5in \vbox{ \epsfxsize=3.0in \epsfbox[0 0 4096 4096]{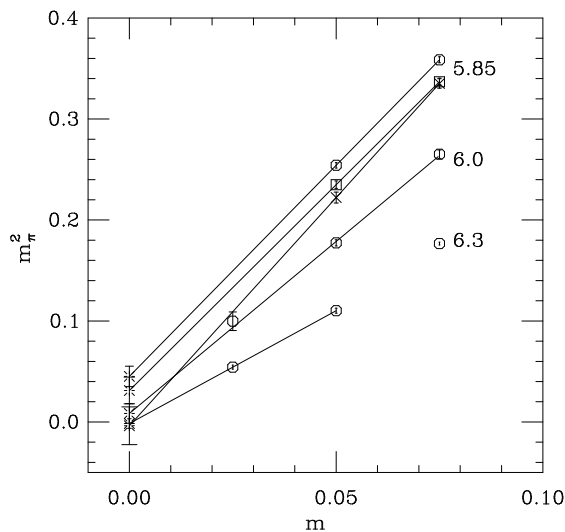} }
    \caption{The pion mass squared. $N_s=10$ (octagons), 14 (squares), 18 (crosses).}
    \label{mpi2}
\end{figure}

In Fig.~\ref{mpi2} we show the pion mass squared as a function of $m$. 
For $N_s=10$ the data at $\gbeta=6.0$ and 6.3 are consistent with 
chiral perturbation theory. However, at $\gbeta=5.85$, $m_\pi^2$ 
extrapolates to a positive non-zero value in  the chiral limit
for $N_s=10$ and 14. There is a large downward shift in the line
as $N_s$ goes from 10 to 14, but it still does not pass through the origin.
However, at $N_s=18$, $m_\pi^2$ extrapolates to -0.004(19) at $m=0$.
The anomalous contribution on the r.h.s. of Eq.~\ref{ratio}
drops from 0.0098(5) to 0.0038(2) as $N_s$ varies from 10 to 18.
It is interesting to note that at $N_s=10$ the anomalous piece
is more than double the value at $\gbeta=6.0$ whereas
the value at $N_s=18$ is roughly the same. When the anomalous term
is sufficiently small compared to the bare parameter $m$, the chiral
symmetry is effectively restored. From the above, it seems we must have
$\LL J_{5q}|\pi\RR/\LL J_5|\pi\RR \simle 0.1 \, m$. 
Of course, as $m\to0$, one must
also take $N_s$ larger,  which is analogous to the situation
with the ordinary spatial volume. Fewer sites 
in the extra dimension may be sufficient at 5.85 if $M$ is increased
still further. 
%
%
%

Research supported by US DOE grant
DE-AC0276CH0016. The numerical computations were carried out on the 
NERSC T3E.

\end{document}